\documentclass[letterpaper]{ptephy}
\usepackage{natbib}
\begin{document}

\title{Inclusive spectrum of the $d(\pi^+, K^+)$ reaction at 1.69 GeV/$c$}

%\author{\name{\fname{First} \surname{Author}}{1}}
%\author{First Author}
\author{\name{\fname{Yudai}~\surname{Ichikawa}}{1, 2,*}, \name{\fname{Tomofumi}~\surname{Nagae}}{1}, \name{\fname{Hyoungchan}~\surname{Bhang}}{3}, \name{\fname{Stefania}~\surname{Bufalino}}{4}, \name{\fname{Hiroyuki}~\surname{Ekawa}}{1, 2}, \name{\fname{Petr}~\surname{Evtoukhovitch}}{5}, \name{\fname{Alessandro}~\surname{Feliciello}}{4}, \name{\fname{Hiroyuki}~\surname{Fujioka}}{1}, \name{\fname{Shoichi}~\surname{Hasegawa}}{2}, \name{\fname{Shuhei}~\surname{Hayakawa}}{6}, \name{\fname{Ryotaro}~\surname{Honda}}{7}, \name{\fname{Kenji}~\surname{Hosomi}}{2}, \name{\fname{Ken'ichi}~\surname{Imai}}{2}, \name{\fname{Shigeru}~\surname{Ishimoto}}{8}, \name{\fname{Changwoo}~\surname{Joo}}{3}, \name{\fname{Shunsuke}~\surname{Kanatsuki}}{1}, \name{\fname{Ryuta}~\surname{Kiuchi}}{8}, \name{\fname{Takeshi}~\surname{Koike}}{7}, \name{\fname{Harphool}~\surname{Kumawat}}{9}, \name{\fname{Yuki}~\surname{Matsumoto}}{7}, \name{\fname{Koji}~\surname{Miwa}}{7}, \name{\fname{Manabu}~\surname{Moritsu}}{10},  \name{\fname{Megumi}~\surname{Naruki}}{1}, \name{\fname{Masayuki}~\surname{Niiyama}}{1}, \name{\fname{Yuki}~\surname{Nozawa}}{1}, \name{\fname{Ryosuke}~\surname{Ota}}{6},  \name{\fname{Atsushi}~\surname{Sakaguchi}}{6}, \name{\fname{Hiroyuki}~\surname{Sako}}{2}, \name{\fname{Valentin}~\surname{Samoilov}}{5}, \name{\fname{Susumu}~\surname{Sato}}{2}, \name{\fname{Kotaro}~\surname{Shirotori}}{10}, \name{\fname{Hitoshi}~\surname{Sugimura}}{2}, \name{\fname{Shoji}~\surname{Suzuki}}{8}, \name{\fname{Toshiyuki}~\surname{Takahashi}}{8}, \name{\fname{Tomonori}~\surname{N.~Takahashi}}{10}, \name{\fname{Hirokazu}~\surname{Tamura}}{7}, \name{\fname{Toshiyuki}~\surname{Tanaka}}{6}, \name{\fname{Kiyoshi}~\surname{Tanida}}{3}, \name{\fname{Atsushi}~\surname{O. Tokiyasu}}{10}, \name{\fname{Zviadi}~\surname{Tsamalaidze}}{5}, \name{\fname{Bidyut}~\surname{Roy}}{9}, \name{\fname{Mifuyu}~\surname{Ukai}}{7}, \name{\fname{Takeshi}~\surname{O. Yamamoto}}{7} and \name{\fname{Seongbae}~\surname{Yang}}{3}}

%\address{First author address}
\address{\affil{1}{Department of Physics, Kyoto University, Kyoto 606-8502, Japan}
    	      \affil{2}{ASRC, Japan Atomic Energy Agency, Ibaraki 319-1195, Japan}
             \affil{3}{Department of Physics and Astronomy, Seoul National University, Seoul 151-747, Korea}
             \affil{4}{INFN, Istituto Nazionale di Fisica Nucleare, Sez. di Torino, I-10125 Torino, Italy}
             \affil{5}{Joint Institute for Nuclear Research, Dubna, Moscow Region 141980, Russia}             
             \affil{6}{Department of Physics, Osaka University, Toyonaka 560-0043, Japan}
             \affil{7}{Department of Physics, Tohoku University, Sendai 980-8578, Japan}
             \affil{8}{High Energy Accelerator Research Organization (KEK),  Tsukuba, 305-0801, Japan}
             \affil{9}{Nuclear Physics Division, Bhabha Atomic Research Centre, Mumbai, India}
             \affil{10}{Research Center for Nuclear Physics, Osaka 567-0047, Japan}
             \email{yudai@scphys.kyoto-u.ac.jp}}

\begin{abstract}%
We have measured an inclusive missing-mass spectrum of the $d(\pi^+, K^+)$ reaction
at the pion incident momentum of 1.69 GeV/$c$ at the laboratory scattering angles
between 2$^\circ$ and 16$^\circ$ with the missing-mass resolution of 2.7~$\pm$~0.1~MeV/$c^2$~(FWHM) 
%referee
at the missing mass of 2.27~GeV/$c^{2}$.
In this Letter, we first try to understand the spectrum as a simple quasi-free picture
based on several known elementary cross sections, considering the neutron/proton
Fermi motion in deuteron. While gross spectrum structures are well understood in this
picture, we have observed two distinct deviations; one peculiar enhancement at
2.13 GeV/$c^2$ is due to the $\Sigma N$ cusp,
and the other notable feature is a shift of a broad bump structure,
mainly originating from hyperon resonance productions of $\Lambda(1405)$ and $\Sigma(1385)^{+/0}$, 
by about %22.4~$\pm$~0.6~MeV/$c^2$ 
22.4~$\pm$~0.4~(stat.)~$^{+2.7}_{-1.7}$~(syst.)~MeV/$c^2$ toward the low-mass side,
%referee
 which is calculated in the kinematics of a proton at rest as the target.
\end{abstract}

\subjectindex{Hyperon production, Kaonic nuclei, $\Lambda(1405)$, Strangeness physics}

\maketitle
\begin{sloppypar}

\section{1. Introduction}\label{sec:intro}

Studies of hyperon productions off nuclei %near the binding thresholds
have provided us useful information on $YN$ interactions.
For example, the $(\pi^+,K^+)$ reaction at 1.05~GeV/$c$ has been used to determine
the $\Lambda$ potential depth in nuclear matter by observing
single-particle energy levels of $\Lambda$ bound states in heavy nuclei~\cite{Hotchi}.
The $\Sigma$-nucleus potential is known to be repulsive in medium-heavy
nuclei~\cite{Saha} from the spectrum shape of the $(\pi^-, K^+)$ reaction
at 1.2~GeV/$c$ near the binding threshold.
Further, importance of $\Sigma N$-$\Lambda N$ coupling was
recognized from the observations of a prominent cusp structure at the $\Sigma N$ threshold 
in $K^- + d \rightarrow \pi^- + \Lambda + p$ reaction~\cite{Braun}.
Recently, the $\Sigma N$ cusp has been intensively investigated in
the $p + p \rightarrow p + K^+ + \Lambda$ reaction at COSY~\cite{COSY-HIRES,COSY-TOF}.
These results were listed in Ref.~\cite{Cusp_summary}.
%Here, the shape of this cusp are not consistent with each other data set and, thus, a theoretical interpretation 
%is difficult. 
%Therefore, further detailed analyses with new data would reveal the information on the $\Sigma N$-$\Lambda N$
%coupling strength and the pole position.

In the present measurement of the $(\pi^+,K^+)$ reaction at~1.69 GeV/$c$,
the productions of hyperon resonances such as $\Lambda$(1405) and
$\Sigma$(1385) are possible. Information on the $\Lambda$(1405)/$\Sigma$(1385)-$N$
interaction would be extracted from the data.
%The measurement was carried out at the K1.8 beam line~\cite{Takahashi} of the 
%J-PARC hadron experimental hall.
%A typical beam intensity was $3\times 10^6$ per
%6-seconds spill cycle with a spill length of about 2 seconds.
%A liquid deuteron target of 1.99-g/cm$^2$ thickness was used for the
%$d(\pi^+, K^+)$ reaction at 1.69 GeV/$c$, and a liquid hydrogen target of 0.85-g/cm$^2$
%thickness for the $p(\pi^+, K^+)$ reactions at 1.58 and 1.69 GeV/$c$ as a calibration.

A goal of the present measurement (J-PARC E27 experiment) is to look for a $K^-pp$, 
a bound state of a $K^-$ with two protons, in the $d(\pi^+, K^+)$ reaction.
In this reaction, one possible production mechanism for the $K^-pp$
was discussed in Ref.~\cite{AY_E27}; a hyperon resonance $\Lambda(1405)$ is produced in the
$\pi^++$``$n$"$\rightarrow K^+ + \Lambda(1405)$ process as a doorway to the $K^-pp$ 
formation through $\Lambda(1405) + $``$p$"$ \rightarrow K^-pp$. Here, ``$n$" and ``$p$" indicate a neutron
and a proton in a deuteron, respectively.
In this respect, it should be noted that the $\pi^- + p \rightarrow K^0 + \Lambda(1405)$ reaction at 
1.69 GeV/$c$ is one of a few reactions, in which $\Lambda(1405)$ was clearly observed ~\cite{Thomas}. %and the 
%mass and width were obtained~\cite{Thomas}. % as 1405~MeV/$c^2$ and 45$\mbox{--}$55 MeV, respectively~\cite{Thomas}.
The incident $\pi^+$ momentum in the E27 was fixed at the same value.

However, a sticking probability of $\Lambda(1405)$ to form the $K^-pp$ would be as low as $\sim$1\%~\cite{AY_E27}.
Thus we need coincidence of some decay products of the $K^-pp$ 
to enhance the signal-to-background ratio.
In the E27 experiment, we installed range counter arrays (RCA)
surrounding the liquid deuterium target for detection of high-momentum 
($\gtrsim$250 MeV/$c$) protons.
The detail of the RCAs and their analyses are described elsewhere. %~\cite{E27coincidence}. 
In this Letter, we focus on the inclusive missing-mass spectrum in
the $d(\pi^+, K^+)$ reaction, and aim to understand the predominant
quasi-free processes in this reaction.
%background processes in this reaction first.

At the pion incident momentum of 1.69 GeV/$c$, a simple
quasi-free picture or an impulse approximation is expected to work
properly~\cite{Fujii}.
Here, we considered the following processes. %in this frame work.
%We construct such an event generator based on differential cross sections for known elementary
%processes taking account of the Fermi motion in deuteron.
%Here we use a deuteron wave-function derived from the Bonn potential~\cite{Bonn}.
In the $d(\pi^+, K^+)$ reaction, not only quasi-free hyperon productions of $\Lambda$ and $\Sigma^{+/0}$,
%\begin{eqnarray}
\begin{equation}
\pi^+ + ``n" \rightarrow K^+ + \Lambda ,\label{eqn:QFlambda} 
\end{equation}
\begin{equation}
%\pi^+ + ``N" \rightarrow K^+ + \Sigma ,\label{eqn:SigmaPlus}
\pi^+ + ``n" \rightarrow K^+ + \Sigma^0;\quad \pi^+ + ``p" \rightarrow K^+ + \Sigma^+ ,\label{eqn:SigmaPlus}
\end{equation}
%\end{eqnarray}
but also hyperon resonance productions of $\Lambda(1405)$ and $\Sigma(1385)^{+/0}$,
\begin{equation}
%\begin{eqnarray}
\pi^+ + ``n" \rightarrow K^+ + \Lambda(1405) ,\label{eqn:LambdaStar}
\end{equation}
\begin{equation}
%\pi^+ + ``N" \rightarrow K^+ + \Sigma(1385) ,\label{eqn:SigmaStar}
\pi^+ + ``n" \rightarrow K^+ + \Sigma(1385)^0;\quad \pi^+ + ``p" \rightarrow K^+ + \Sigma(1385)^+ ,\label{eqn:SigmaStar}
%\end{eqnarray}
\end{equation}
and non-resonant productions of $\Lambda \pi$ and $\Sigma \pi$, 
\begin{equation}
%\pi^+ + ``N"\rightarrow K^+ + Y + \pi ,\label{eqn:PhaseSpace}
\pi^+ + ``N"\rightarrow K^+ + \Lambda + \pi;\quad \pi^+ + ``N"\rightarrow K^+ + \Sigma + \pi, \label{eqn:PhaseSpace}
\end{equation}
take place.
The cross sections of elementary processes have been already measured~\cite{Thomas, Pan}.

There are two previous measurements on the $d(\pi^+, K^+)$ reaction in this momentum region.
One measurement is at a lower incident momentum of 1.4 GeV/$c$~\cite{Pigot}, 
%so that hyperon resonance productions were not observed.
in which the production of hyperon resonances, $\Sigma(1385)$ and $\Lambda(1405)$, is kinematically forbidden.
On the other hand, the $\Sigma N$ cusp was observed in the missing-mass spectrum around 2.13~GeV/$c^2$
in the events with a large multiplicity measured by counters surrounding the target.
The other measurement using a deuterium bubble chamber was carried out at incident beam momenta
between 1.1 and 2.4 GeV/$c$~\cite{Davies}.
They measured incident energy dependence of total cross sections for several reactions as well as
the $\Lambda$-$\pi^+$ invariant-mass distributions in the $\pi^+ + d \rightarrow K^+ + \Lambda + \pi^+ + (n_{s})$ reaction.
%$\pi^+ + p \rightarrow K^+ + \Lambda + \pi^+$ reaction
%mode from the invariant mass of $\Lambda + \pi^+$.
%Thus, 
The present measurement is the first measurement
of the inclusive $d(\pi^+, K^+)$ missing-mass spectrum covering a wide missing-mass
region from $\Lambda$ and $\Sigma$ to $\Lambda(1405)/\Sigma(1385)$.

\if0
The other measurement was carried out with almost same incident momentum of 1.7 GeV/$c$~\cite{Davies}
by using a deuterium bubble chamber. They observed the $\Sigma(1385)$ production
in $\pi^+ + ``p" \rightarrow K^+ + \Lambda + \pi^+$ with a neutron as a spectator
as well as several other reaction modes.
The present measurement is the first measurement
of the inclusive $d(\pi^+, K^+)$ missing-mass spectrum covering a wide missing-mass
region from $\Lambda$ and $\Sigma$ to $\Lambda(1405)/\Sigma(1385)$.
\fi

\section{2. Experimental Setup}
The measurement was carried out at the K1.8 beam line~\cite{Takahashi} of the 
J-PARC hadron experimental hall.
%The detail informations of the K1.8 beam line and the experimental setup of this measurement were 
%shown in Ref.\cite{Takahashi, YI_FB}.
The detailed information was described in Ref.~\cite{Takahashi, YI_FB}.
A typical beam intensity was $3\times 10^6$ per
6-second spill cycle with a spill length of about 2~seconds.

The incident pion beams were measured with the K1.8 beam line spectrometer.
Their momenta were reconstructed by using a third-order transfer matrix with
the %expected 
estimated momentum resolution of 0.18 $\pm$ 0.01$\%$~(FWHM).

A liquid deuterium target of 1.99-g/cm$^2$ thickness was used for the
$d(\pi^+, K^+)$ reaction at 1.69 GeV/$c$, and a liquid hydrogen target of 0.85-g/cm$^2$
thickness for the $p(\pi^+, K^+)$ reactions at 1.58 and 1.69 GeV/$c$ for calibrations.
%We used a liquid target system, which contained cylindrical-shaped target cell,  cooled with liquid helium. 
The target shape was cylindrical; 67.3~mm in diameter and 120~mm in length.
The target fully covered the incident beam with a typical beam size of 
$\sigma_{H} = 7.6$~mm (horizontal) and $\sigma_{V} = 4.2$~mm (vertical).

Scattered kaons were analyzed with the Superconducting Kaon Spectrometer (SKS).
The particle momentum  was obtained with the Runge-Kutta method by using a calculated field map. 
%Tracks were reconstructed with the Runge-Kutta method by using a calculated field map and the particle momenta 
%were estimated from these values. 
The SKS magnetic field was set at 2.36~T. Emitted particles in the momentum range of 0.8$\mbox{--}$1.3 GeV/$c$ with
a scattering angle between 2$^{\circ}$ and 16$^{\circ}$ were analyzed. 
%referee
%The very forward events (less than 2$^{\circ}$) were cut out in order to suppress the beam backgrounds. 
The very forward events (less than 2$^{\circ}$) were cut out 
in order to keep good vertex resolution and suppress beam-oriented backgrounds in the $(\pi^+,K^+)$ events,
such as $\mu^+$ from $\pi^+$ decay and 
secondary particles produced from various support structures. 
%We cut out less than 2$^{\circ}$ in order to suppress the beam backgrounds.
%%%%%%%%%%%%%%%%%%%%%%%%%%%%%%%%%%%%%%%%%%%%%%%%%%%%%%%%%%%%%%%%%%%
By using SKS, the particle momenta were measured with a resolution of about 0.2$\%$~(FWHM) in an acceptance range 
of about 100~msr.
Kaons were identified by a time-of-flight in combination with the flight path length
obtained for each track.

\begin{figure}
\begin{center}
\vspace*{-0.5cm}
\includegraphics[height=55mm]{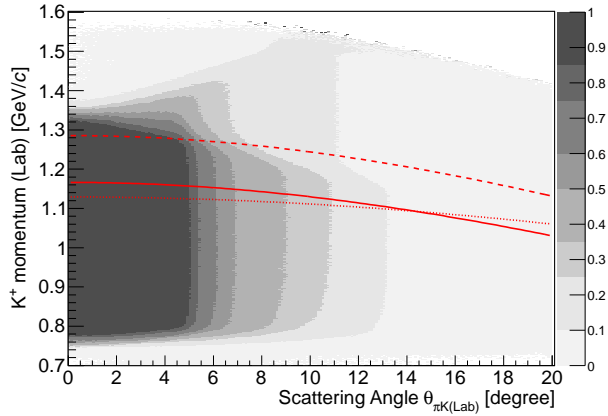}
\vspace*{-0.4cm} 
 \caption{Acceptance of SKS. %The horizontal (vertical) axis is scattering angle (momentum) in Lab. system.
Three kinematical lines of $K^+$ are %momentum and scattering angle
also shown  for the $\pi^+ + p\rightarrow K^+ + \Sigma^+$ reactions 
at 1.58 GeV/$c$~(solid~line) and 1.69~GeV/$c$~(dashed~line) and
the $\pi^+ + d \rightarrow K^+ + [K^-pp]$ reaction at 1.69~GeV/$c$ assuming the binding energy of 100 MeV~(dotted~line).} 
\label{fig:Acceptance}
\end{center}
\end{figure}

The acceptance of SKS is shown in Fig. {\ref{fig:Acceptance}} as a function of the 
particle momentum and scattering angle. 
It was obtained with a Monte Carlo simulation based on Geant4~\cite{geant}.
%taking account of beam distributions.
In the figure, three kinematical lines %three kinematical lines of $K^+$ momentum and scattering angle 
are shown  for the $p(\pi^+, K^+)\Sigma^+$ reactions
% $\pi^+ + p \rightarrow K^+ + \Sigma^+$ reactions 
at 1.58~GeV/$c$~(solid~line) and 1.69~GeV/$c$~(dashed line) and the $d(\pi^+, K^+)K^-pp$ reaction at 
%$\pi^+ + d \rightarrow K^+ + [K^-pp]$ reaction at
1.69~GeV/$c$ assuming the binding energy of 100~MeV~(dotted~line).
%Note that the kinematical lines of the first reaction at 1.58 GeV/$c$ and the $d(\pi^+, K^+)K^-pp$
%have a large overlap with the almost flat momentum-acceptance region.

%
%
% Analysis
\section{3. Analysis}
After selecting a $(\pi^+, K^+)$ event requiring a fiducial volume cut
in the target, a missing-mass spectrum is obtained as a double differential 
cross section of $d^{2}\bar{\sigma}/d\Omega/dM$ 
%$\displaystyle \frac{d^2\bar{\sigma}}{d\Omega dM}$ 
% $d^{2}\bar{\sigma}/d\Omega/dM$ 
averaged over the scattering angle from 2$^{\circ}$ to 16$^{\circ}$. 
%in a unit of $\mu$b/sr/(MeV/$c^2$). 
The double differential cross section is calculated as
\begin{equation}\label{eq:eq2}
\frac{d^2\bar{\sigma}}{d\Omega dM} = \frac{A}{N_{A}(\rho x)} \frac{N_{K}}{N_{beam}\Delta \Omega \Delta M \epsilon},
\end{equation}
where $A$ is the target mass number, $N_{A}$ the Avogadro constant, $\rho x$ the target mass thickness, %g/cm$^2$ 
$N_{K}$ the number of detected kaons in the missing-mass interval $\Delta M$, $N_{beam}$ the number of beam pions on the target, 
$\Delta \Omega$ the solid angle of SKS and $\epsilon$ 
the overall experimental efficiency resulting from DAQ, detectors, and analysis cuts
including the $K^+$ decay factor.
A typical value of the experimental efficiency is $\epsilon = 17.5 \pm 0.9 \%$. 
%and the total systematic uncertainty caused by the ambiguity of these experimental efficiency on the differential
%cross section is estimated to be 5$\%$.

%\subsection{3.1. Calibrations}
\section{3.1. Calibrations}
First we need to adjust the missing-mass scale of the $d(\pi^+, K^+)$ reaction. 
The relative momentum scale between the beam line spectrometer and 
SKS was studied by introducing a beam at 0.9 GeV/$c$ through the
two spectrometers without a target.
Further, we can study the missing-mass scale in the
$p(\pi^+, K^+)$ reaction at 1.58 and 1.69 GeV/$c$ by looking at the $\Sigma^+$ mass.
By combining all these information, the incident beam momentum
was corrected with a linear function as $p^{\rm{corr}}_{\rm{K1.8}} = 1.0077 \times p_{\rm{K1.8}}-0.0087$ GeV/$c$, 
where $p^{\rm{corr}}_{\rm{K1.8}}$ is the corrected beam momentum and $p_{\rm{K1.8}}$ is the measured beam momentum. 
The systematic uncertainty of the momentum scale is
estimated to be $\pm$1.1 MeV/$c$ from these calibrations.
%was corrected with a linear function of the measured beam momentum ($p_{K1.8}$) as 
%$p_{corrK1.8} = 1.0077 p_{K1.8}-0.0087$, $p_{corrK1.8}$ is the corrected beam momentum in unit of GeV/$c$.

We also optimized the missing-mass resolution of 
$p(\pi^+, K^+)\Sigma^+$ reaction at 1.58~GeV/$c$
by correcting the $K^+$ momentum with 
a fifth order polynomial in the horizontal ($dx/dz$) and vertical ($dy/dz$) direction cosines
of the particle orbit.
After this correction, the missing-mass resolution 
was obtained to be 2.8 $\pm$ 0.1~MeV/$c^2$~(FWHM), which corresponds to
the missing-mass resolution of 3.2~$\pm$~0.2~MeV/$c^2$~(FWHM)
at the $\Sigma N$ cusp region in the $d(\pi^+, K^+)$ reaction.
This resolution is at least by a factor of 2 better than the previous measurement with
the same reaction (3.2$\mbox{--}$5.5 MeV/$c^2$ in $\sigma$)~\cite{Pigot}.
%This value is about a half of the previous measurement (3.2$\mbox{--}$5.5 MeV/$c^2$ in $\sigma$)~\cite{Pigot}
%for the cusp.

\begin{figure}
\vspace*{-0.5cm} 
 \begin{minipage}{0.5\hsize}
  \begin{center}
 \hspace*{-0.4cm}
       \includegraphics[width=70mm]{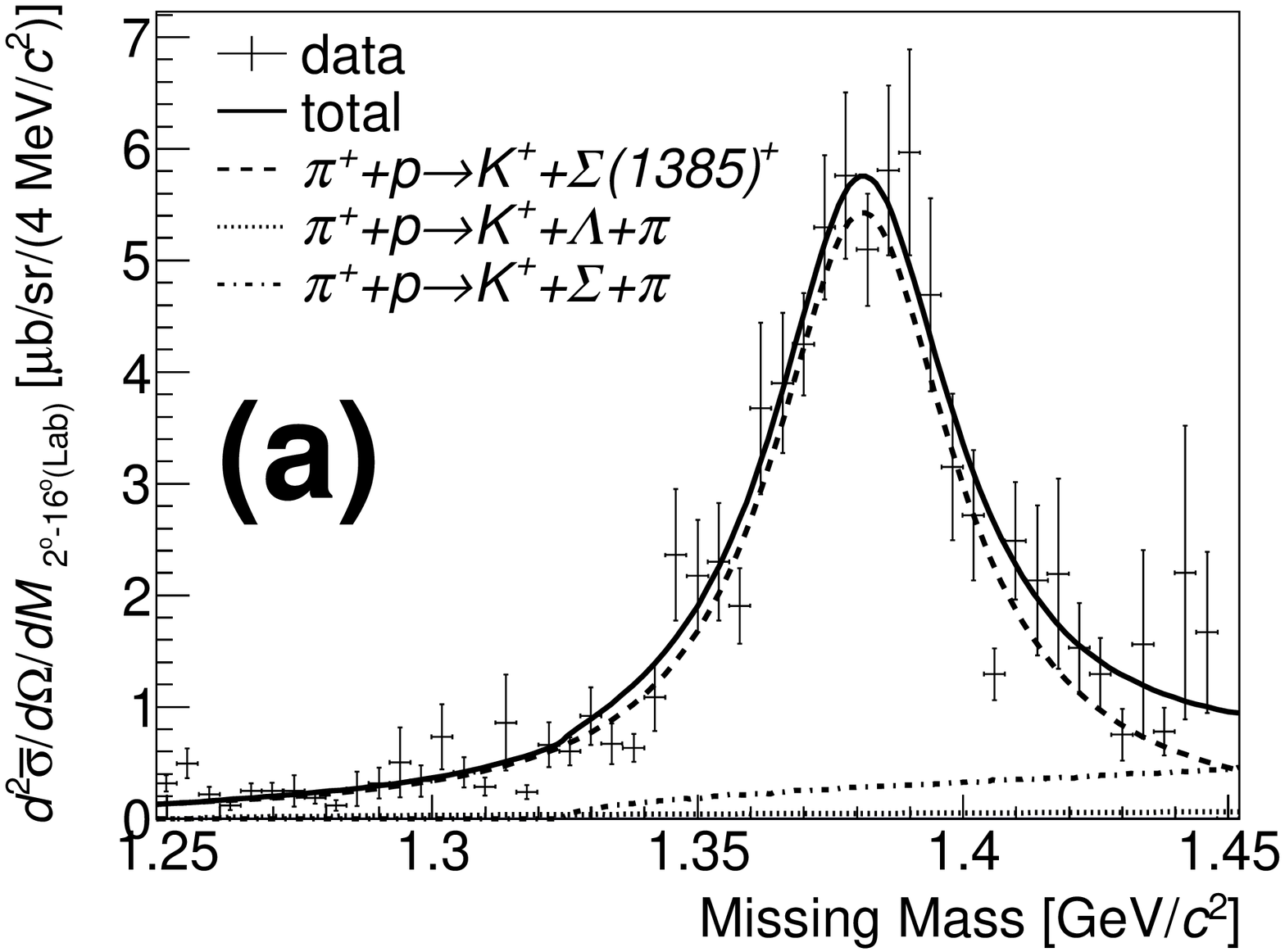} 
  \end{center}
 % \vspace*{-0.7cm}  
 \end{minipage}
%\hspace*{0.3cm} 
 \begin{minipage}{0.5\hsize}
  \begin{center}
  \vspace*{-0.5cm}
     \includegraphics[width=70mm]{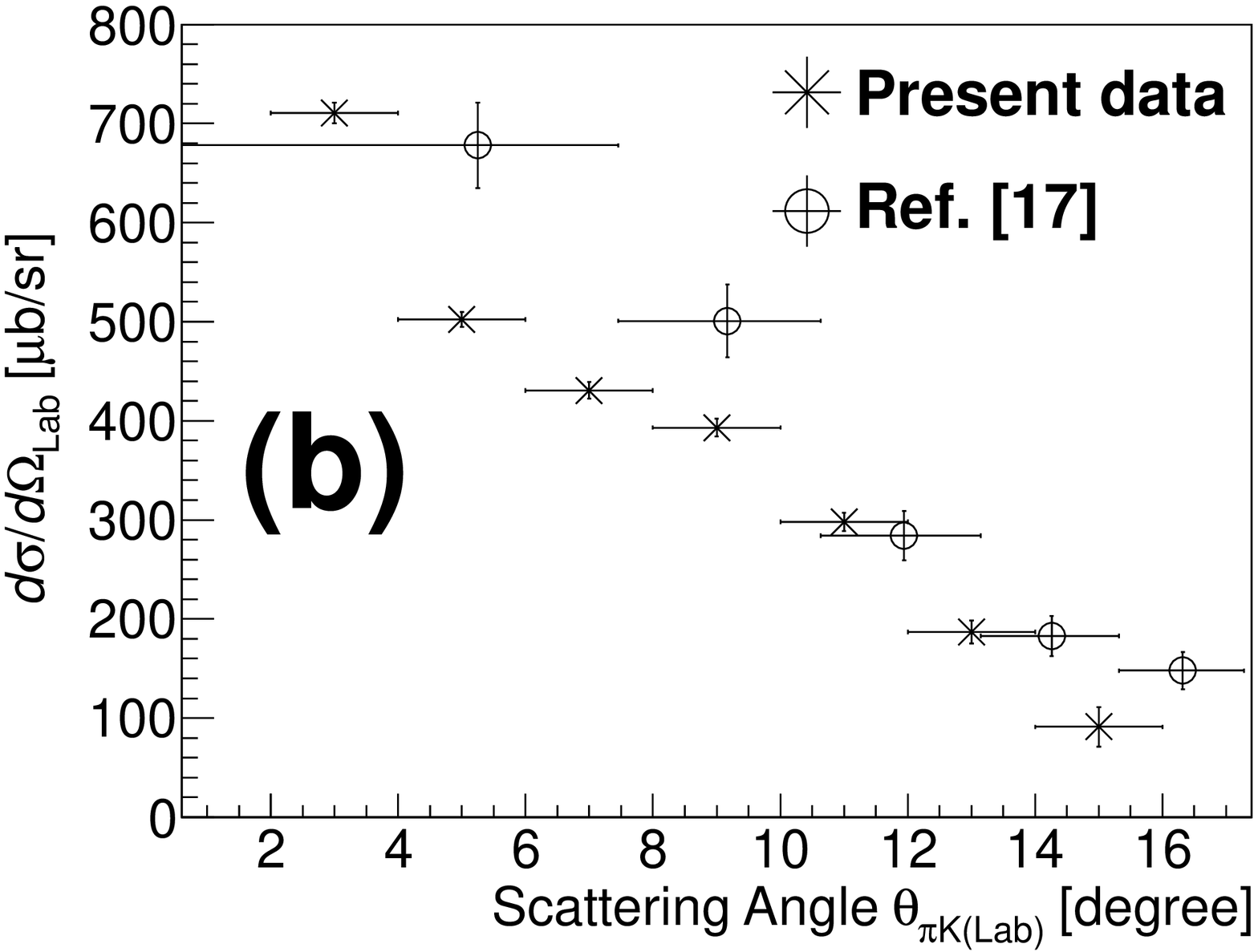}  
  \end{center} 
 \vspace*{-0.7cm} 
  \end{minipage}
 \caption{(a) The missing-mass spectrum of the $p(\pi^+, K^+)$ reaction in the $\Sigma^+(1385)$ mass region. 
     The experimental data are shown by black points with statistical errors. 
     The spectrum was fitted with $\Sigma(1385)^+$~(dashed line), $\Lambda \pi$~(dotted line) and $\Sigma \pi$~(dot-dashed line) productions. 
  (b) The differential cross sections of $\Sigma^+$ production at $p_{\pi^+}$~=~1.69~GeV/$c$. The present data and the referenced old ones are shown by crosses with statistical errors and open circles, respectively.}
\label{fig:Calibrations}
\end{figure}

After the missing-mass scale adjustment,
the validity of the corrections was tested in the
$\pi^+ + p\rightarrow K^+ + \Sigma(1385)^+$ productions at 1.69 GeV/$c$.
%We have obtained the $\Sigma^+$ mass at $1189\pm 0.04$ MeV/$c^2$,
%which is consistent with the PDG value.
In Fig.~\ref{fig:Calibrations} (a), we show the missing-mass spectrum in the
$\Sigma(1385)^+$ region in the $p(\pi^+,K^+)$ reaction at 1.69 GeV/$c$.
A fitting result with a Lorentzian function for the $\Sigma(1385)^+$~(dashed line)
and the three-body phase space distributions for the $\Lambda \pi K^+$~(dotted line) and 
the $\Sigma \pi K^+$~(dot-dashed line) is also shown.
The $\Sigma(1385)^+$ mass and width are found to be 1381.1~$\pm$~3.6~(stat.)~MeV/$c^2$ 
and 42~$\pm$~13~(stat.)~MeV, respectively, which are consistent with the
PDG values~\cite{PDG} within the statistical errors.
%The systematic uncertainty of the momentum scale is
%estimated to be $\pm$1.1 MeV/$c$ from the results of the beam through data at 0.9 GeV/$c$, 
%the $p(\pi^+, K^+)\Sigma^+$ data at 1.58 and 1.69 GeV/$c$ and the $p(\pi^+, K^+)\Sigma(1385)^+$
%data at 1.69 GeV/$c$.
%Considering this momentum scale uncertainty, the missing mass scale uncertainty on
%the $K^-pp$ is estimated to be 0.7 MeV/$c^2$.

%Next, we check the validity of our cross section estimations 
%by comparing the $\Sigma^+$ production cross sections
%in the $p(\pi^+, K^+)$ reactions at 1.58 and 1.69 GeV/$c$
%with old measurements.
Next, we compared the cross section of the $p(\pi^+,K^+)\Sigma^+$ reaction between our measurement and an old measurement with the same momenta of 1.58 and 1.69 GeV/c by Candlin~{\it et al.}~\cite{Candlin}.
As shown in Fig.~\ref{fig:Acceptance}, the kinematical
lines at two different incident momenta run through 
different acceptance areas. 
%referee %%%%%%
%In both cases, we obtained a good agreement of  our data 
%with the previous ones.
In both cases, we obtained a reasonable agreement between our data and the previous ones within the large errors.
%%%%%%%%%%%%%%%%%%%%%%%%%%%%
Here we show the case for 1.69 GeV/$c$ in Fig.~\ref{fig:Calibrations} (b).
%referee %%%%%%
%The obtained differential cross section of $\Sigma(1385)^+$ was 
%also consistent with the past experiment~\cite{Pan}.
The same was true in the cross section of $\Sigma(1385)^{+}$~\cite{Pan}.
%%%%%%%%%%%%%%%%%%%%%%%%%%%%%%%%%%%%%%%%%%%%%%%%%%%%%%%%%%%%

%% figure : cross section of Sigma+

\section{3.2. $\pi^+ + d \to K^+ + X$ at 1.69 GeV/$c$}
\begin{figure}
\vspace*{-0.5cm} 
  \begin{minipage}{0.5\hsize}
  \begin{center}
\vspace*{0.05cm} 
\hspace*{-0.4cm}
\includegraphics[width=70mm]{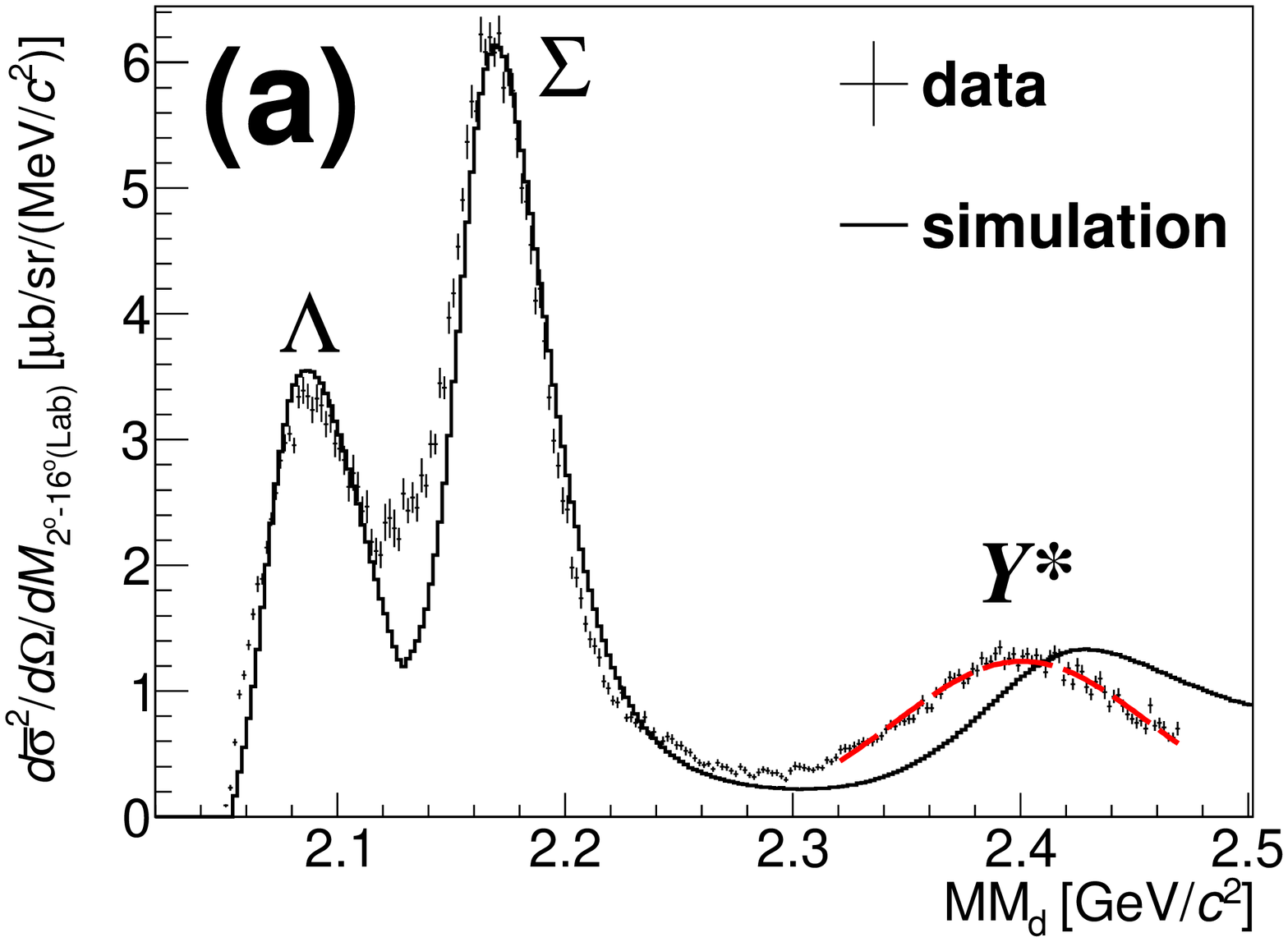}
 \end{center} 
% \vspace*{-0.7cm} 
 % \caption{The differential cross sections of $\Sigma^+$ production at $p_{\pi^+}$ = 1.69 GeV/$c$. The present data and the referenced old data are shown by red and blue crosses, respectively.}
\end{minipage}
\begin{minipage}{0.5\hsize}
  \begin{center}
 \hspace*{-0.4cm}
 \includegraphics[width=70mm]{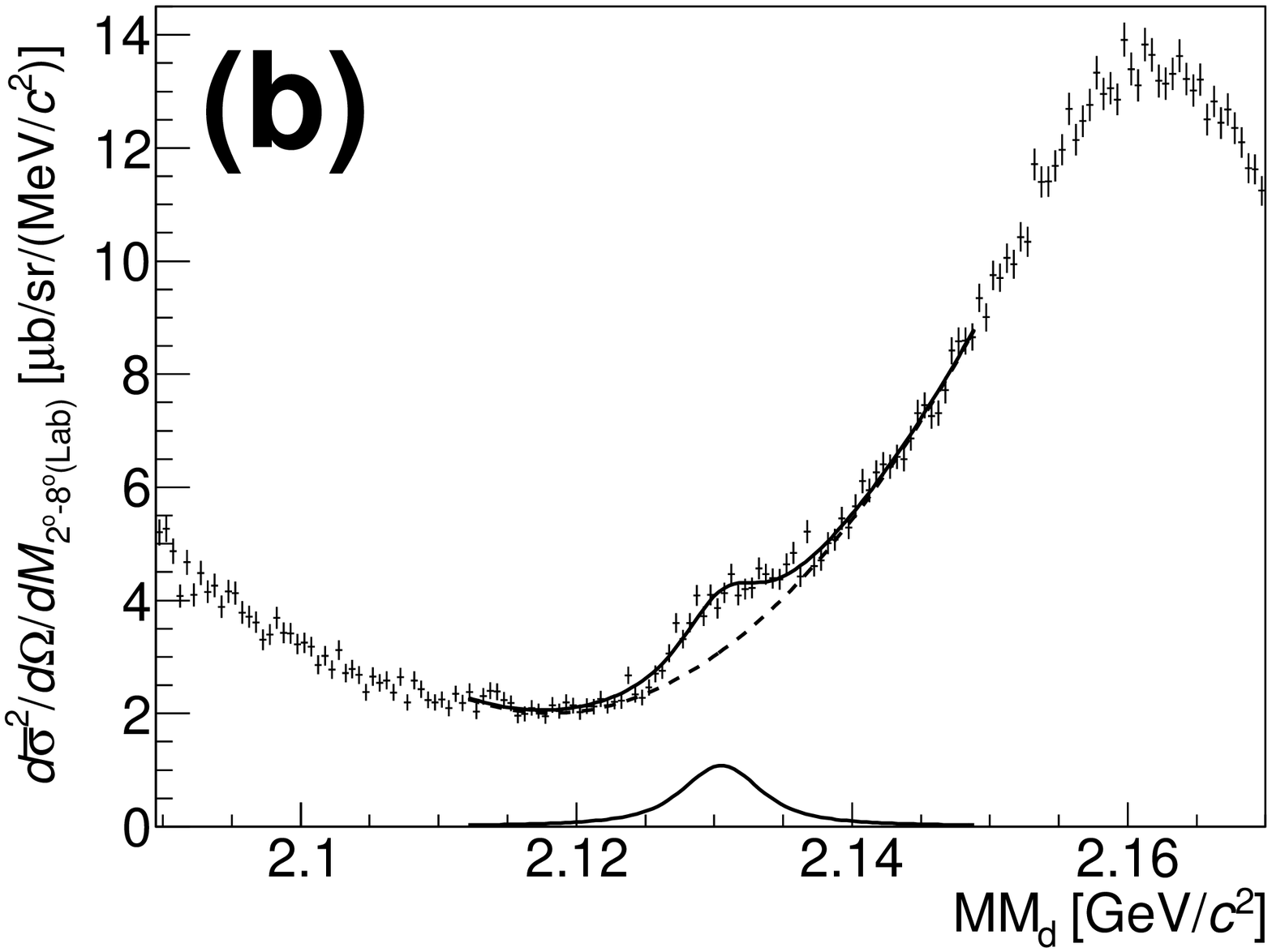}
  \end{center}
  %\vspace*{-0.7cm} 
  %\caption{The obtained missing mass for $\pi^+p\to XK^+$ reaction at $p_{\pi^+}$~=1.69~GeV/$c$.}
  %\label{fig:MM_P}
 \end{minipage}
\vspace*{-0.3cm} 
\caption{
(a) The missing-mass spectrum ($MM_{d}$) of the $d(\pi^+, K^+)$ reaction for the scattering angle from 2$^{\circ}$ to 16$^{\circ}$(Lab) per 2 MeV/$c^2$.
%which is calculated deuteron as a target per 2 MeV/$c^2$. 
The crosses and solid line show the experimental data and the simulated spectrum, respectively. The result of the $Y^*$ peak fitting is also shown with a dashed red line for the experimental data.
(b) The missing-mass spectrum (MM$_{d}$) in 2.09 to 2.17 GeV/$c^2$ region for the forward scattering angle from 2$^{\circ}$ to 8$^{\circ}$(Lab) per 0.5 MeV/$c^2$, which is shown by crosses. The fitting results are shown by solid and dashed lines ($\chi^2/ndf$ = 1.11). See details in the text.}
\label{fig:MM_d}
\end{figure}

%Figure~\ref{fig:MM_d} (Left) shows the obtained missing-mass spectrum for the $d(\pi^+, K^+)$ reaction at
%1.69 GeV/$c$ where the missing-mass $MM_d$ is calculated  with a deuteron as the target.
Figure~\ref{fig:MM_d} (a) shows the obtained missing-mass (denoted as $MM_{d}$) spectrum for the $d(\pi^+, K^+)$ reaction at 
1.69 GeV/$c$. %where the missing-mass $MM_d$ is calculated  with a deuteron as the target.
As we discussed in Sec.~1, 
we can find three major structures in 
the spectrum: quasi-free $\Lambda$ component from the reaction (\ref{eqn:QFlambda}), 
quasi-free $\Sigma$ component from the reaction (\ref{eqn:SigmaPlus}), and 
quasi-free $Y^*$ component from the reactions  (\ref{eqn:LambdaStar}), (\ref{eqn:SigmaStar}). 
The non-resonant phase space component from the reaction (\ref{eqn:PhaseSpace}) 
constitutes a broad structure under the quasi-free $Y^*$ bump.

We made an attempt to reproduce the double differential cross section $d^{2}\bar{\sigma}/d\Omega/dM$
%$\displaystyle %\frac{d^2\bar{\sigma}}{d\Omega dM}$ 
with a simulation by using the cross sections $d\sigma/d\Omega$ 
%$\displaystyle \frac{d\bar{\sigma}}{d\Omega}$ 
of each reaction
obtained in the past experiments with a smearing by the nucleon Fermi motion in a deuteron.
Here, we used a deuteron wave-function derived from the Bonn potential~\cite{Bonn}.
%We used the present data for the quasi-free $\Sigma^+$ component.
%Others were used as summarized in Table~\ref{table:cross_sections}.
For the $\pi^++``n"$ reactions, we used the cross sections and angular distributions of $\pi^- + p$ reactions in Ref.~\cite{Thomas} assuming the isospin symmetry.
For the $\pi^+ + ``p"$ reactions, we used the values in Ref.~\cite{Pan}.

Here, since the cross section for the quasi-free $\Sigma^0$ process~\cite{Thomas,Sigma0_1,Sigma0_2} features
rather large errors in the forward angles, an adjustment for the normalization
of the cross section was applied within the measurement errors for the quasi-free $\Sigma$ component.
%Others are listed in Table~\ref{table:cross_sections}.
By taking into account the Fermi motion $p_F$, the participant nucleon is assumed to be in the
off-mass shell as $M_p^{*2}= \left(M_d-\sqrt{M_{s}^2+p_F^2}\right)^2-p_F^2$, where $M_d$ and $M_s$
are deuteron and the spectator on-shell nucleon mass %, while the other spectator nucleon is in the on-mass shell
(spectator model~\cite{COSY-TOF-SM}).
The outgoing $K^+$ momentum, $p_{K}$, was distributed according to the reaction kinematics  
with the mass of the participant nucleon $M_p^*$ and its momentum $\vec{p_F}$. 
%following to the reaction kinematics.
Then, the missing-mass $MM_d$ was calculated as $MM_d^2=(E_\pi+M_d-E_K)^2-|\vec{p_\pi}-\vec{p_K}|^2$.
Thus, we obtained the simulation result as shown in Fig.~\ref{fig:MM_d}~(a) indicated by a solid line. 

%
% Missing-mass spectra

We find an overall structure of the spectrum is well reproduced except for two distinct differences.
One difference is a cusp structure observed at around 2.13 GeV/$c^2$; a magnified view is shown 
in Fig.~\ref{fig:MM_d}~(b) for the forward scattering angles from 2$^{\circ}$ to 8$^{\circ}$ in the laboratory frame. 
A peak at $\Sigma N$ thresholds (2.1289 GeV/$c^2$ for $\Sigma^+ n$ and 2.1309 GeV/$c^2$ for $\Sigma^0 p$) 
is prominent in the figure. 
%referee
When we chose the scattering angle larger than 8$^{\circ}$, the cusp is less prominent due to the large quasi-free backgrounds. 
%%%%%%%%%%%%%%%%%%%%%%%%%%%%%%%
Although this  $\Sigma N$ cusp may not necessarily distribute according to a Lorentzian function, 
here we fit the cusp structure with this function 
in order to compare with the previous results summarized in Ref.~\cite{Cusp_summary}.
%here we fit the cusp structure with a Voigt function which expresses a Lorentzian function folded with a Gaussian
%in order to compare with the previous results summarized in Ref.~\cite{Cusp_summary}.
Through a fit of the Lorentzian function folded with the resolution of 1.4~MeV/$c^{2}$ in $\sigma$ for the cusp~(solid line) and a
third-order polynomial function for a continuum background~(dashed line), 
we obtained the peak position at $2130.5~\pm~0.4$~(stat.)~$\pm$~0.9~(syst.)~MeV/$c^2$,
the width of $\Gamma$~=~5.3~$^{+1.4}_{-1.2}$~(stat.)~$^{+0.6}_{-0.3}$~(syst.)~MeV
and the differential cross section of $d\bar{\sigma}/d\Omega$ = 10.7~$\pm$ 1.7 $\rm{\mu b/sr}$.
The $\chi^2/ndf$ of this fitting was 1.11.
The systematic errors of these values were estimated in $\sigma$ taking into account uncertainties 
in the missing-mass scale, fitting ranges, the missing-mass resolution ($\pm$ 0.08~MeV/$c^{2}$), 
the binning of the missing-mass spectrum and background functional shapes by changing the third to fifth order polynomials. 
This result is the first observation of the $\Sigma N$ cusp structure in the inclusive spectrum of the $d(\pi^+,K^+)$ reaction.
%, etc.
%background functional shapes by changing the 3rd to 5th order polynomials, etc.

Such a cusp can appear at the opening of a new threshold in order to conserve the flux and the associated unitarity of the $S$-matrix. 
However, the cusps are not always seen in experimental cross sections.  
%as an enhancement in a 
On the other hand, a cusp structure can be pronounced when a pole exists near the threshold~\cite{Badalian}. 
Miyagawa and Yamamura~\cite{Miyagawa} suggested that the poles exist near the $\Sigma N$ threshold in a second or third quadrant of the complex plane of the $\Sigma N$ relative momentum by using several $YN$ potential models. 
Therefore, the cusp structure at the $\Sigma N$ threshold would not be a 
simple threshold effect but could be caused by a nearby pole.

The obtained peak position is consistent with the previous measurements in Ref.~\cite{Cusp_summary}. 
In several reactions existence of a shoulder at about 10 MeV higher mass was reported~\cite{Cusp_summary}. 
%referee
We can conclude nothing on the existence of the shoulder structure because of the large quasi-free $\Sigma$ backgrounds.
%We cannot observe this structure in our spectrum because of the large quasi-free $\Sigma$ backgrounds. 
In addition, the width seems to be smaller than the averaged value of 12.2~$\pm$~1.3~MeV in other reactions~\cite{Cusp_summary}. 
%the averaged value of other reactions was 12.2~$\pm$~1.3~MeV. 
%In addition, the width seems to be smaller than the values obtained in other reactions~\cite{Cusp_summary}, 
%the averaged value of other reactions was 12.2~$\pm$~1.3~MeV. 
In order to discuss the possible pole position, 
we need realistic theoretical calculations taking into account the $(\pi^+, K^+)$ reaction mechanism. 
It will be very interesting to further compare our results including angular distributions and 
%referee
%coincidence analyses with such calculations.
coincidence data with high momentum protons in the RCA. 
%%%%%%%%%%%%%%%%%%%%%%%%%%%%%%%%%%%%%%%%%%%%%%%%%%%%%%%%% 
Note that we can suppress the quasi-free $\Lambda$/$\Sigma$ productions in the coincidence data because the decay particles of such hyperons are out of the RCA acceptance.

The other difference is a ``shift" of the broad bump position for the $Y^*$ productions. 
In this region the contributions of $\Sigma(1385)^{+/0}$ and $\Lambda(1405)$ overlap each other, 
and it is not possible to disentangle them in an inclusive measurement. 
When we fit the bump with a Gaussian function, we obtained the
peak position at  2400.6~$\pm$~0.5~(stat.)~$\pm$~0.6~(syst.)~MeV/c$^2$ for the present data
and at 2433.0~$^{+2.8}_{-1.6}$~(syst.)~MeV/$c^2$ for the simulation. 
The systematic error for the simulation was estimated taking into account uncertainties of the differential cross sections of $Y^*$, the $Y^*$ mass and fitting ranges.  
The same fitting procedure was applied for a $MM_{p}$ ($= \sqrt{(E_\pi+M_p-E_K)^2-|\vec{p_\pi}-\vec{p_K}|^2}$) spectrum,  which is the missing-mass spectrum calculated
assuming a proton at rest as the target. %proton target kinematics,
We obtained the peak position 
at 1376.1~$\pm$~0.4~(stat.) $\pm$ 0.5~(syst.)~MeV/$c^2$ and 1398.5~$^{+2.6}_{-1.6}$~(syst.)~MeV/$c^2$, respectively.
The amount of the peak shift is 32.4~$\pm$~0.5~(stat.)~$^{+2.9}_{-1.7}$~(syst.)  (22.4~$\pm$~0.4~(stat.)~$^{+2.7}_{-1.7}$~(syst.))~MeV/$c^2$ for the $MM_d$ ($MM_p$) spectra to the low mass side.
Even if the peak position of the simulated spectrum was adopted instead of the fitted one, the difference is reduced by just 4~MeV/$c^2$. 
There could be an uncertainty of several MeV to the low-mass side arising from our simple spectator model in the simulation, 
which was estimated from the difference of the peak positions when the balance of the off-shellness 
between the participant and spectator nucleons was changed. 
%referee
The small difference of about 3~MeV/$c^{2}$ for the quasi-free $\Sigma$ productions is within this uncertainty. 
%The missing-mass spectrum might be also affected by the possible final state interaction between $\Sigma$ and $N$.
%The small difference of about 3~MeV/$c^{2}$ for the $\Sigma$ productions can be understood within this uncertainty, 
%although the missing-mass spectrum can be affected by the final state interaction between $\Sigma$ and $N$. 
%%%%%%%%%%%%%%%%%%%%%%%%%%%%%%%%%%%%%%%%%%%%%%%%%%

More sophisticated theoretical analyses taking into account $Y^{*} N$ interactions in the final state might explain the observed puzzling ``shift".

It should be noted that the LEPS group recently reported a similar inclusive spectrum for the 
$d(\gamma, K^+ \pi^-)$ reaction at a 1.5\mbox{--}2.4 GeV photon energy region and 
found no significant shift in the $Y^*$ region~\cite{Tokiyasu}. 
As we mentioned in Sec.~1, the old deuterium bubble chamber experiment measured 
the invariant mass distribution of $\Lambda + \pi^+$ in the $\pi^+ + d \rightarrow K^+ + \Lambda + \pi^+ + (n_{s})$ reaction. 
The $\Sigma(1385)^+$ mass and width in deuteron turned out to be 
1386.6~$\pm$~4.4~MeV/$c^2$ and 49~$\pm$~11~MeV when it decays into $\Lambda + \pi^+$. %at 1.7 GeV/$c$ pion incident momentum when it decays into $\Lambda + \pi^+$.

\section{4. Summary}
The inclusive missing-mass spectrum of the $d(\pi^+, K^+)$ reaction at the beam momentum of 1.69 GeV/$c$
was obtained in high statistics and high energy resolution for the first time.
The present data cover a wide missing-mass range from the $\Lambda$ production
threshold to the $\Lambda(1405)/\Sigma(1385)$ region.
The overall structure of the spectrum was understood with a simple quasi-free
picture based on the known elementary processes. 

However, there were two peculiar deviations from this picture.
One observation is the $\Sigma N$ cusp, of which 
position was found to be $2130.5~\pm 0.4$~(stat.)~$\pm~0.9$~(syst.)~MeV/$c^2$
with the width of $\Gamma$ =  5.3~$^{+1.4}_{-1.2}$~(stat.)~$^{+0.6}_{-0.3}$~(syst.)~MeV. 
The peak position is consistent with previous measurements. %and the width is smaller than the previous ones.
%This difference might be come from the reaction mechanism. 
%Therefore, we need realistic theoretical calculation of this reaction to deepen understanding. 
Further detailed studies including the present data would reveal the information on the $\Sigma N$-$\Lambda N$ coupling strength and the pole position.
Moreover, the centroid of the broad bump structure in the $Y^*$ production region
was significantly shifted to low mass side as compared with a simple quasi-free simulation,
by about  32.4~$\pm$~0.5~(stat.)~$^{+2.9}_{-1.7}$~(syst.)  (22.4~$\pm$~0.4~(stat.)~$^{+2.7}_{-1.7}$~(syst.))~MeV/$c^2$ for $MM_d$ ($MM_p$) spectra.
In order to clarify the origin of the peak shift, further experimental and theoretical studies are necessary. 

\end{sloppypar}

\ack
%We would like to acknowledge the great efforts of the  Hadron beam channel groups  
%and J-PARC accelerator groups to improve the beam quality and 
%operate the accelerator stably. 
%We also thank Professors T. Yamazaki, Y. Akaishi, T. Harada and K. Miyagawa
We would like to thank the Hadron beam channel group, 
accelerator group and cryogenics section in J-PARC for their great efforts on
stable machine operation and beam quality improvements.
The authors thank the support of NII for SINET4.
We also wish to acknowledge valuable discussions with
Prof.~T.~Yamazaki, Prof.~Y.~Akaishi, Prof.~T.~Harada and Prof.~K.~Miyagawa.
This work was supported by the Grant-In-Aid for Scientific
Research on Priority Area No. 449 (No. 17070005), 
the Grant-In-Aid for Scientific Research on Innovative Area
No. 2104 (No. 221055506), from the Ministry of Education,
Culture, Sports, Science and Technology (MEXT) Japan, and 
Basic Research (Young Researcher) No. 2010-0004752 from 
National Research Foundation in Korea.
%This work was supported by the Grant-In-Aid for Scientific
%Research on Priority Areas No. 449
%and on Innovative Areas No. 2104.
%This work was partly supported by the Grant-in-Aid for Scientific Research on Innovative Areas 
%(No. 22105506) from the Ministry of Education, Culture, Sports, Science and Technology (MEXT), Japan.
We thank supports from National Research Foundation, WCU program of the
Ministry of Education, Science and Technology (Korea), Center for Korean
J-PARC Users.

\bibliographystyle{ptephy}
%\bibliographystyle{plain.bst}
%\bibliographystyle{abbrv}
%\bibliographystyle{authordate1}
%\bibliography{sample}

\end{document}